\documentclass[letterpaper]{article}
\usepackage{aaai20}
\usepackage{times}
\usepackage{helvet}
\usepackage{courier}
\usepackage[toc,page]{appendix}
\usepackage{hyperref}
\usepackage{url}
\usepackage{calligra}
\usepackage{comment}
\usepackage{bm}
\usepackage[neverdecrease]{paralist}
\usepackage{graphicx}
\usepackage{booktabs}
\usepackage{graphics}
\usepackage{tabularx}
\usepackage{xspace}
\usepackage{graphicx}
\usepackage{booktabs}
\usepackage{mdwlist}
\usepackage{multirow}
\usepackage{amsmath}
\usepackage{mathtools}
\usepackage{mathrsfs}
\usepackage{enumitem}
\usepackage{calc}
\usepackage[neverdecrease]{paralist}
\usepackage[ruled,vlined]{algorithm2e}
\usepackage[round]{natbib}

\usepackage{booktabs}
\usepackage{diagbox}
\usepackage{colortbl}
\usepackage{subcaption}
\usepackage{makecell}
\usepackage{soul}
\usepackage{amsfonts}

\renewcommand{\vec}[1]{\mathbf{#1}}

\newcommand{\vecxj}{\vec{x}_j}
\newcommand{\y}{y(\vec{x})}
\newcommand{\ith}{$i^{\mathrm{th}}$}

\newcommand{\STWR}{`Bias-Agnstc-L2'~}
\newcommand{\STWOR}{`Bias-Agnstc'~}
\newcommand{\STWRG}{`Bias-Agnstc-Gndr-Ftr'~}
\newcommand{\STWRR}{`Bias-Agnstc-Race-Ftr'~}
\newcommand{\STWRDW}{`Bias-Agnstc-DWE'~}
\newcommand{\MTLG}{`Bias-Awr-Gndr'~}
\newcommand{\MTLR}{`Bias-Awr-Race'~}
\newcommand{\MTLB}{`Bias-Awr-Joint'~}

\begin{document}

\title{Towards Socially Responsible AI: Cognitive Bias-Aware Multi-Objective Learning
}

\author{Procheta Sen\textsuperscript{\rm 1}, Debasis Ganguly\textsuperscript{\rm 2} \\ 
ADAPT Centre, Dublin City University, Ireland\textsuperscript{\rm 1}, IBM Research, Dublin, Ireland.\textsuperscript{\rm 2} \\
procheta.sen@adaptcentre.ie\textsuperscript{\rm 1}, debasis.ganguly1.ie.ibm.com\textsuperscript{\rm 2}
}


\maketitle

\begin{abstract}
Human society had a long history of suffering from cognitive biases leading to social prejudices and mass injustice.
The prevalent existence of cognitive biases in large volumes of historical data can pose a threat of being manifested as unethical and seemingly inhumane predictions as outputs of AI systems trained on such data.
To alleviate this problem, we propose a bias-aware multi-objective learning framework that given a set of identity attributes (e.g. gender, ethnicity etc.) and a subset of sensitive categories of the possible classes of prediction outputs, learns to reduce the frequency of predicting certain combinations of them, e.g. predicting stereotypes such as `most blacks use abusive language', or `fear is a virtue of women'. Our experiments conducted on an emotion prediction task with balanced class priors shows that a set of baseline bias-agnostic models exhibit cognitive biases with respect to gender, such as women are prone to be afraid whereas men are more prone to be angry. In contrast, our proposed bias-aware multi-objective learning methodology is shown to reduce such biases in the  predictid emotions.
\end{abstract}

\section{Introduction}

Throughout the course of history, human society has witnessed the `us vs. them' conflict,
where certain sections of the society have exhibited bias (in the form of hatred, prejudices, repression and even violence) towards other communities.
This social discrimination has manifested itself in different forms, identified by a wide range of different characteristics, such as gender, ethnicity, religion and caste, among others. Some devastating consequences of social discrimination in history include reducing a woman to a mere legal possession of her husband in ancient Greece \citep{Blundel}, depriving admission to a deserving black student in a South-African university \citep{Schaefer}, or preventing a boy of a Hindu priest family to marry a girl of a lower caste in an Indian village \citep{RameshChandra}, among many others. 

\begin{figure}[t]
    \centering
    \includegraphics[width=.75\columnwidth]{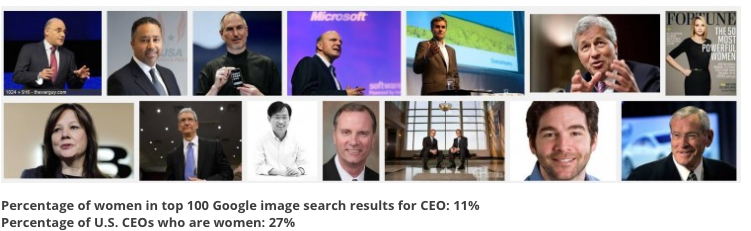}
    \includegraphics[width=.24\columnwidth]{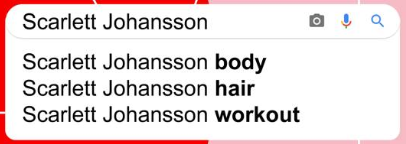}
    \caption{Examples of social bias in existing AI systems: `Google Image Search' under-representing women as CEOs (left), and `Google query completion' prioritizing \emph{appearance} over \emph{filmography} for a popular female actor (right).}
    \label{fig:bias-in-existing-AI-tools}
\end{figure}

Scientific and technological revolution has played a pivotal role in mitigating social biases and prejudices to a large extent in modern human society. Yet the advent of data-driven AI poses a great threat to shift the equilibrium again because these data-driven AI-based predictions are prone to essentially \emph{pick up} the cognitive biases and prejudices from content that was authored during the yester-years. To corroborate this point,
existing studies have shown that word embedding reflects the gender stereotypes present in large volumes of text \citep{Bolukbasi:2016,manzini-etal-2019}, e.g. \emph{men are more likely to be computer programmers while women are more likely to be home-makers}.
To cite further examples, existing studies on prominent existence of social bias in commercially deployed AI systems report that,
\begin{inparaenum}[a)]
\item gender proportions in `Google Image search' are exaggerated according to career-related stereotypes images retrieved in response to the query `CEO' \citep{Kay:2015} (left of Figure \ref{fig:bias-in-existing-AI-tools});
\item  `Google query completion' suggests queries related to the filmography of a male actor, whereas for a female one it typically focuses on her appearance \citep{JamesTemperton2019} (right of Figure \ref{fig:bias-in-existing-AI-tools}); and
\item a recidivism AI system deployed by a number of US states (falsely) predicts high risks for black people \citep{ProPublica}. 
\end{inparaenum}

Developing AI systems that appear to be more \emph{human-like} in the responses they generate \citep{Scheutz2007,Cambria:2016} can lead to the risk of
associating a human-like persona to AI systems,
which in turn can lead to the disastrous consequence of a section of the society into (falsely) \emph{believing} that the biased responses (such as the ones illustrated in Figure \ref{fig:bias-in-existing-AI-tools}) are also human in nature.

A possibility in feature-based models is to manually intervene and leave out the features that could lead to biased predictions for a task, e.g. the New York Police Department (NYPD) refrains from using the \emph{race} of a person to predict the risks of future crimes \citep{ProPublica}. However, understanding which features or their combinations could lead to \emph{ethically correct} responses for a particular task is not always easy, unless such a bias resurfaces out from the data and is actually observed, e.g., it is difficult to see what features (term weighting functions) could implicitly lead to the observation about female actors in Figure \ref{fig:bias-in-existing-AI-tools} (right), which is nothing short of `body shaming'. This situation is obviously aggravated for data-driven neural models, which rely on learning an abstract representation of the data, with an associated risk that this abstract representation is likely to be considerably different from how humans would `abstractify' the data themselves for the purpose of generating not just correct but \emph{ethically correct} responses.
%

Different from these existing studies on debiasing,
which can be broadly classified into the ones that remove bias from either a) embedded words, or b) language models or c) training data itself,
we propose a multi-task learning based approach to achieve a trade-off between the \emph{correctness} and \emph{fairness} of a model. Specifically, we propose a generic approach to first quantify and then reduce bias \emph{jointly} against a number of secondary \emph{social identity} attributes (e.g. gender and ethnicity) in a \emph{classification} task involving a single \emph{primary} attribute (e.g. predicting emotions, such as fear, anger etc. from a piece of text). Specifically, we employ a multi-objective learning approach, where the intention is to increase the social acceptability (fairness) of the predicted outputs without causing a significant degradation in the effectiveness of the primary classification task (correctness).

\section{Related Work}

\paragraph{Demonstrating social bias in classification tasks.}

\citet{davidson-etal-2019} demonstrated racial bias in four different tweet data sets used for hate and abusive language detection. They showed that a classifier trained on these tweets
exhibits a social bias in predicting that abusive tweets are
mostly written by African-Americans as compared to white Americans.
On a similar note, \citet{sap-etal-2019} demonstrated racial bias in Twitter datasets for abusive language detection. 
They showed that when presented with the ethnicity information, the annotators were less likely to tag a tweet of an African American as abusive compared to the situation when annotators were not present with the racial bias information. 
\citet{kiritchenko-mohammad-2018-examining} reports that a number of existing models for sentiment analysis suffer from either race or gender bias.
\citet{Zafar:2017} used a constrained learning approach to debias classification models on feature vectors.
In contrast, we employ a multi-objective function and do not use hand-crafted features for our experiments.

\paragraph{Social identity features for improving predictions.}

\citet{hovy-2015-demographic} showed that the use of demographic features, such as age, and gender can improve the text classification task performance across five languages. On a similar note, \citet{vanmassenhove-etal} found that including the gender information of the speakers (authors) help to translate sentences more effectively
to a target language with gender-specific morphology.
\citet{stanovsky-etal-2019} proposed an evaluation methodology to measure gender bias in machine translation (MT) systems and also released a dataset to support further investigation in MT bias.
\citet{garimella-etal-2019} released a gender tagged dataset for POS-tagging and dependency parsing and showed that POS-tagging and dependency parsing effectiveness can vary across genders.


\paragraph{Mitigating bias from word embedding and down-stream tasks.}

\citet{prost-etal-2019} proposed ways to debias word embedding with an objective to reduce bias in downstream tasks.
\citet{Bolukbasi:2016} debiased pre-trained embedded word vectors
obtain gender-neutral word vectors, e.g. in such an embedded space `babysit' is expected to be equidistant from grandmother and grandfather.
Given an existing word embedding, their approach involves
defining a gender specific subspace and then
learning a linear transformation which seeks to preserve pairwise inner products between the word vectors while minimizing the projection of the gender neutral words onto the gender subspace.
%
\citet{zhao-etal-2018} proposed another approach to obtain debiased  word embedding by preserving the gender information with additional dimensions, the presence of non-zero components along which indicates gender inclined words.

\citet{gonen-goldberg-2019} concludes that existing word embedding debiasing techniques (e.g. \cite{Bolukbasi:2016})
are rather superficial in nature, in the sense that applying such debiased embeddings can still lead to gender bias in down-stream tasks. \citet{manzini-etal-2019} proposed a multi-class debiasing solution for word embedding. They removed race, gender and racial bias from existing word vectors.
\citet{may-etal-2019-measuring} measured social bias in sentence encoders. \citet{zhao-etal-2019} proposed debiasing approaches for contextualized word embeddings by using data augmentation and neutralization.

\citet{qian-etal-2019} proposed a gender debiasing method for language generation by introducing an additional term in the loss function for language generation, seeking to make the probability more uniform across both genders.
\citet{sun-etal-2019} reviews existing gender-bias detection techniques for NLP and the advantages and disadvantages of the debiasing approaches. \citet{zhang-etal-2019} proposed bias detection and debiasing methods for sentence paraphrasing.
\citet{bevendorff-etal-2019} proposed an author verification method that also illustrates the sources of bias in the corresponding corpus. They also showed that elimination of bias sources can lead to a more balanced dataset.

\section{Bias-Aware Predictions}

\paragraph{Primary classification task.}
Before explaining our proposed framework for
multi-objective debiasing, we formalize the problem definition with the following notations.
Let
\begin{equation}
X=\{(\vecxj, y(\vecxj)\}_{j=1}^{M}, \vecxj \in \mathbb{R}^d,
y(\vecxj) \in P=\{0,\ldots,k\}
\end{equation}
be a set of $d$-dimensional real-valued vectors, each with a categorical variable, $y=y(\vec{x})$, 
which are the labels of a primary task with possible values in a range of $k+1$ different categories.

As a concrete example, each $\vec{x}$ could represent an embedded vector representation of a sentence, whereas
the $y(\vec{x})$ values could correspond to $k=2$ categories of emotion, namely `anxiety' and `joy' associated with a given sentence, mapped to integers $\{0,1,2\}$ (0: none).
Generally speaking, the primary task is then to learn a model for predicting a category, $\y$, given an input vector, $\vec{x}$, i.e.,
\begin{equation}
\phi: (\vec{x},\y(\vec{x})) \mapsto \Delta_{k},\,\, \vec{x} \in X \label{eq:primary-transformation},    
\end{equation}
where $\Delta$ represents a $k$-simplex of class probabilities of primary task categories.


\begin{table}[t]
    \centering
    \small
    \begin{tabularx}{\columnwidth}{@{}l@{~~}X@{}}
    \toprule
    $\{(\vec{x}, y(\vec{x}))\}_{j=1}^{M}$ & $M$ pairs of data and primary task labels \\
    $P \in \{0,\ldots,k\}$ & Primary task label categories \\ 
    $P_s \in \{0,\ldots,k_s\}$ & A subset of $k_s (< k)$ categories, (e.g. `fear') \\
    $n$ & \#identity attributes (e.g. `gender') \\
    $z_i$ & Categorical variable for the \ith identity attribute with $m_i$ possible values \\
    $C_i=\{0,\ldots,m_i\}$ & Set of categories of the \ith identity attribute (e.g. `male', `female') \\
    $U_i \subset C_i$ & A set of (historically) under-represented categories, e.g. `female' \\
    $D_i \subset C_i$ & Historically dominating categories, e.g. `male'
    \\
    $Y^B_i \in \{0,1\}$ & Indicator variables for learning the pseudo-task of generating biased responses.\\
    \bottomrule
    \end{tabularx}
    \caption{Summary of notations.}
    \label{tab:notations}
\end{table}

\paragraph{(Social) Identity attributes.}
A model, such as the one represented in Equation \ref{eq:primary-transformation} can lead to socially unacceptable responses. To make ethically correct predictions, it is necessary to remove such cognitive biases. To quantify bias formally, we define a set of categorical attributes, akin to \emph{social identity}, against which a \emph{specific subset} of predicted output categories may exhibit socially unacceptable biases.
There are two important points to take notice of.

First, the set of categorical attributes corresponding to social identities, henceforth referred to as \emph{identity attributes}, is not a part of the input data. This is because explicitly using such features for training a model, $\phi$ (Equation \ref{eq:primary-transformation}), may be socially unacceptable in the first place, e.g., NYPD avoids using the race of a person with criminal records to predict the risk of his/her future crimes \citep{ProPublica}. 

Second, only a specific combination of the identity attributes with a subset of the primary output categories may be deemed as socially unacceptable, (the complementary combinations being less likely to be controversial). As a concrete example, frequently predicting the emotion of `fear' with the identity attribute `female' is disturbing in today's society (rightly so), whereas its complement, i.e. frequent associations of `fear' with `man' is not (possibly attributed to the desire to change a long history of patriarchal society). 

\paragraph{Formulating cognitive bias.}
Next, we introduce a set of secondary response variables $y^B$ to define the association between a subset, say $P_s=\{0,\ldots,k_s\} \subset P$,
of primary categories ($k_s < k$) and a set of $n$ identity attributes, say $\{z_i\}_{i=1}^n$, where each $z_i \in \{0,\ldots,m_i\}$, i.e. is a categorical variable with $m_{i+1}$ possible values, each mapped to an integer in $[0,m_i]$ (e.g. the `Gender' attribute with `male' mapped to 0 and `female' to 1).
For the purpose of defining bias, we now assume that this set of categories $C_i=\{0,\ldots,m_i\}$ corresponding to the identity attribute $z_i$, is comprised of two mutually disjoint subsets, i.e. $C_i=U_i \cup D_i$, where $U_i$ denotes the set of historically under-represented categories with respect to the set $P_s$.

The following concrete example is used to clarify the idea (using words rather than integers for readability). For an identity attribute `gender',
$C_{\mathrm{gender}}=\{\mathrm{male}, \mathrm{female}\}$, if the primary task is to predict emotion from one of $P=\{\mathrm{fear},\mathrm{anger}\}$, then an example of cognitive bias results by defining
$P_s=\{\mathrm{fear}\}$ and $U_{\mathrm{gender}}=\{\mathrm{female}\}$ (women are prone to be more afraid than men).

\begin{figure}[t]
    \centering
    \includegraphics[width=.8\columnwidth]{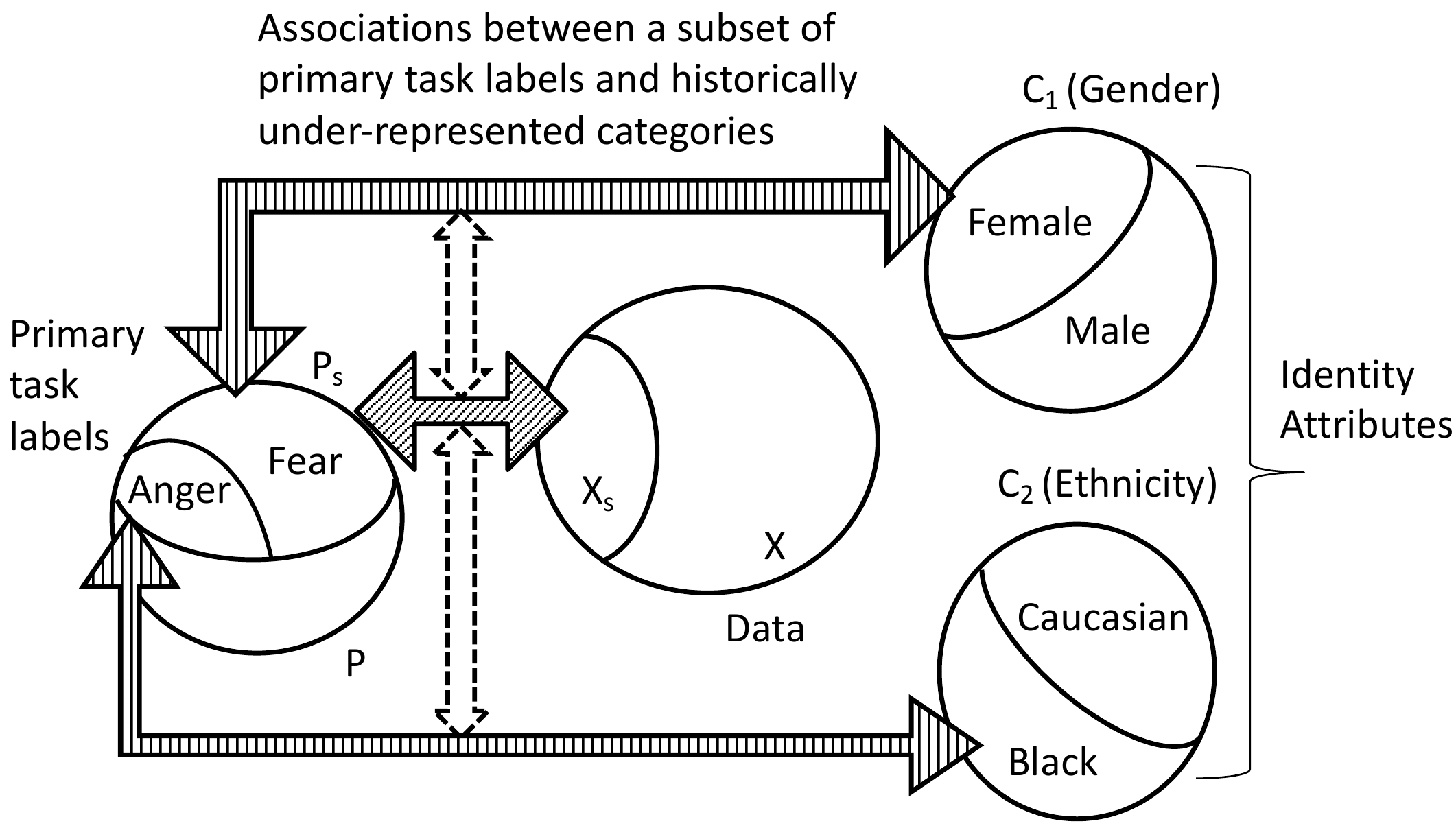}
    \caption{Schematic of cognitive bias removal.}
    \label{fig:schematic-bias}
\end{figure}

\paragraph{Bias response variables.}

We then define a bias response as a Boolean variable denoting a frequent association between a subset of primary categories, $P_s$, and a particular identity attribute $z_i$ as a function of the likelihood of the association. More formally,
\begin{equation}
y^B_i(\vec{x}) =
\begin{cases} 
1, & \frac{\mathbb{I}(y(\vec{x}) \in P_s \land z_i \in U_i)}{\mathbb{I}(y(\vec{x}) \in P_s)} > \tau\\
0, & \mathrm{otherwise}  \label{eq:yb-def}
\end{cases}
\end{equation}
where $\tau\in [0,1]$ is a parameter (set to $\frac{1}{2}$ in all our experiments).
Equation \ref{eq:yb-def} sets the Boolean variable $y^B_i=0$ if a) $y(\vec{x}) \in P-P_s$, i.e., $y(\vec{x})$ is a category which is neutral to a given set of identity attributes, or b) if the co-occurrence of the primary task labels ($P_s$) into a set of specific categories, likely to be associated with social biases, is sufficiently high (determined by a parameter $\tau \in [0,1]$).     
With respect to the schematic shown in Figure \ref{fig:schematic-bias}, two sets of bias variables
are defined for the two identity attributes, namely gender and race (shown towards the right), using the co-occurrences of specific values of these attribute (e.g. female and black) with prediction categories `fear' and `anger' respectively. 


\paragraph{Pseudo-task of biased response generation.}
Using the bias response variables of Equation \ref{eq:yb-def}, the next step is then to learn a function mapping from input vectors, their associated primary task labels along with the corresponding indicator pseudo-variables for bias detection (as per Equation \ref{eq:yb-def}), $(\vec{x}, y, y^B_i)$, to
a Bernoulli probability distribution indicating the probability of presence (or absence) of social bias corresponding to the $i$-th identity attribute.
More formally,
\begin{equation}
\phi^B_i: (\vec{x}, y^B_i) \mapsto \Delta_{1},\,\, \vec{x} \in X_s = \{\vec{x}: y(\vec{x}) \in P_s\}.  \label{eq:biastaskmap}
\end{equation}
Equation \ref{eq:biastaskmap} corresponds to learning the associations between the inputs, primary task labels and an identity attribute; two such relations, one for gender, and the other for ethnicity, are shown with the two vertical dotted arrows in Figure \ref{fig:schematic-bias}.
Although the indicator variables denote the presence of bias, in our learning step we effectively invert the variables so as to intentionally perform poorly for the task of generating biased responses, which eventually leads to decreasing the bias from the primary prediction task.

\begin{figure}[t]
    \centering
    \includegraphics[width=.85\columnwidth]{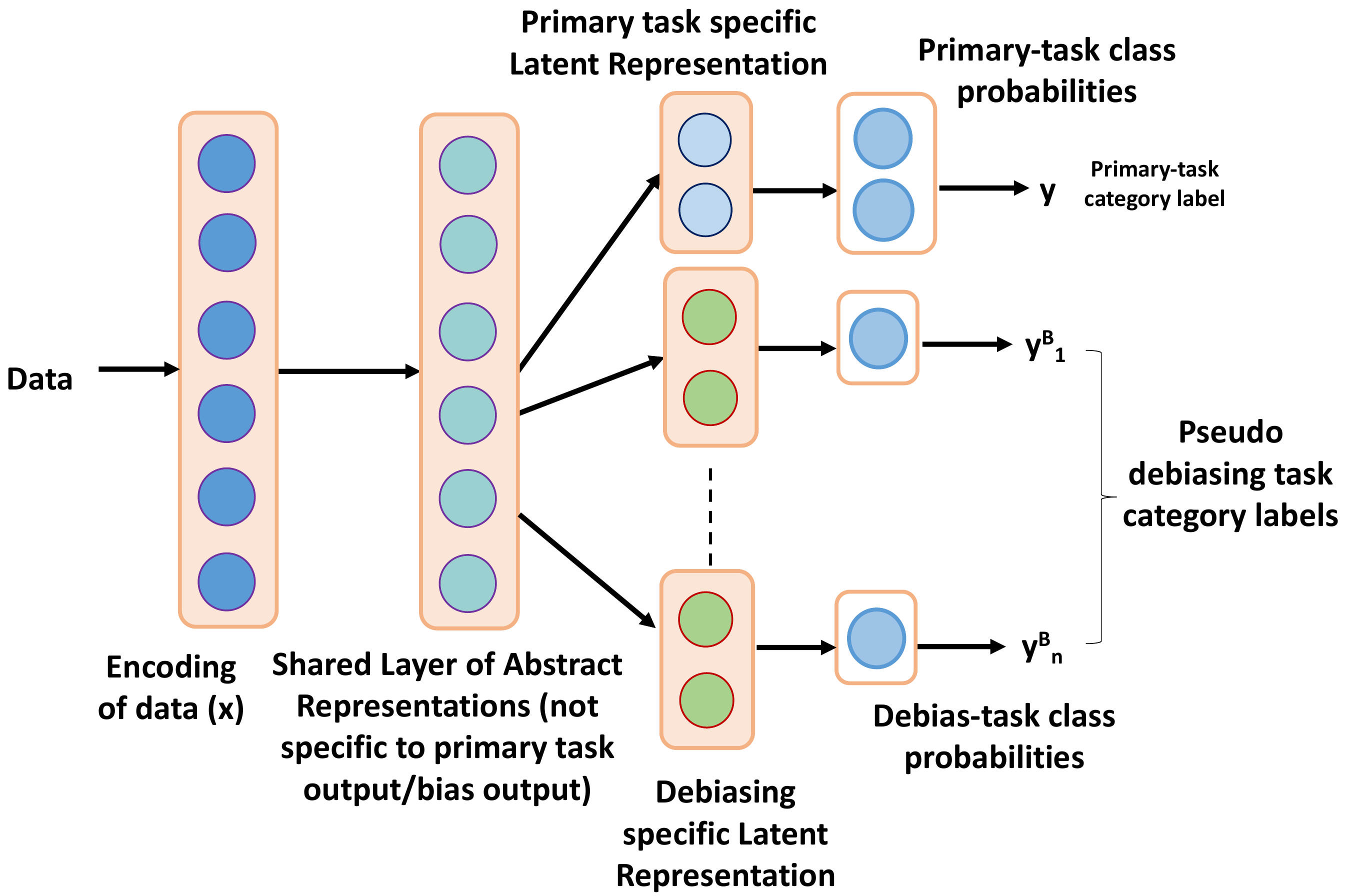}
    \caption{Schematic diagram of a neural network architecture for \emph{jointly} learning the primary task objective (for effective prediction) and a set of debiasing tasks (for reducing the cognitive bias in these predictions).}
\label{fig:deBiasArch}
\end{figure}

\paragraph{Multi-Objective Neural Architecture.}

Next, we learn the
primary classification task (Equation \ref{eq:primary-transformation}) \emph{simultaneously} along with the $n$ different bias tasks, i.e. one for each identity attribute (Equation \ref{eq:biastaskmap}).
This joint learning can specifically be realized with a neural architecture, schematically depicted in Figure \ref{fig:deBiasArch}. Concretely speaking, we first employ a linear transformation, $\Theta_s \in \mathbb{R}^{d\times p}$ to transform each data vector to a \emph{shared} abstract representation, and then apply a set of subsequent linear transformations specific to the primary and the bias tasks, respectively.
\begin{equation}
\begin{split}
\hat{y} & = \mathrm{softmax}(\Theta_p(\Theta_s(\vec{x}))), \Theta_p \in \mathbb{R}^{p \times k} \\
\hat{y^B_i} & = \mathrm{sigmoid}(\Theta^B_i(\Theta_s(\vec{x}))), \Theta^B_{i} \in \mathbb{R}^{p \times 1}, \vec{x} \in X_s \label{eq:softmax}.
\end{split}
\end{equation}
We then maximize the following joint likelihood function to learn the shared layer and the task specific parameters.
\begin{equation}
\mathcal{L}=P(y|\vec{x};\Theta_p,\Theta_s) - \sum_{i=1}^{n} P(y^B_i|\vec{x};\Theta^B_{i},\Theta_s), \label{eq:jointloss}
\end{equation}
where the respective probabilities are estimated from Equation \ref{eq:softmax}. Specifically, for our experiments, we use square loss to back-propagate errors and compute gradients. 

An important observation in Equation \ref{eq:jointloss} is the \emph{negative} sign in the likelihood of the bias tasks, which indicates that the overall objective is to perform well in the primary task and perform \emph{poorly} in each bias pseudo-task seeking to reduce the non-uniformity in the posterior distribution of the respective categories of identity attributes with respect to a subset of primary task categories.
Intuitively speaking, the joint objective seeks to sacrifice a little on the correctness of predictions to be more ethically correct.

\section{Evaluation}

Before describing the main experiments, we start this section with an illustrative example on a two dimensional synthetic dataset, where we visualize the working principle of bias removal our proposed model. We used this synthetic dataset to investigate if the proposed model can be effective on a relatively simple setup, which would then provide evidence that it could also be potentially effective on real data as well.

\subsection{Visualization with Synthetic 2D Data}

\begin{figure}[t]
\centering
\includegraphics[width=0.49\columnwidth]{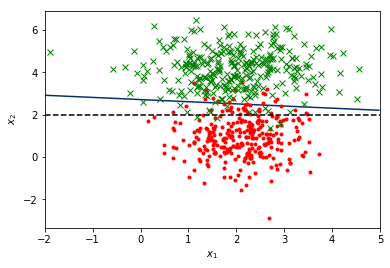}
\includegraphics[width=0.49\columnwidth]{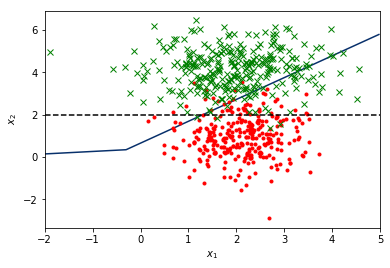}
\caption{Illustrative example to visualize bias reduction with multi-objective learning (Equation \ref{eq:jointloss}). Predictions
with a bias-agnostic classifier (logistic regression with $L_2$ regularization) are effective but exhibits a `bias' in associating the green points with the top half of the plot area (left); whereas the multi-objective learning is able to reduce such a bias (right).
\label{fig:2d-data}}
\end{figure}

Figure \ref{fig:2d-data} plots a set of $2000$ samples, $\vec{x}=(x_1,x_2)$, generated from a two dimensional iso-tropic Gaussians mixture model (GMM) of two components (plotted as red and green, respectively) with equal priors; the centres located at $\mu_1^{T}=(2,1)$ (red) and $\mu_2^T=(2,4)$ (green). To illustrate the idea of bias in predictions, we assume that the set $P_s$ (the set of categories likely to be attached as social stereotypes as per the notations of Table \ref{tab:notations}) in this example corresponds to the category `green', i.e. $P_s=\{2\}$. As an identity attribute, we consider the $x_2$ variable; the set $C_{x_2}$ comprised of points
\begin{equation}
U_{x_2}=\{(x_1,x_2) \in \mathbb{R}^2: x_2 > 2\} \label{eq:2dexample}
\end{equation}
and its complement set $D_{x_2}$.

Assuming that the primary task in this example involves predicting the correct component of a set of points sampled from the GMM (as specified above), we trained our proposed bias-aware predictive model by first substituting $P_s=\{2\}$ and $U_{x_2}$ from Equation \ref{eq:2dexample} into Equation \ref{eq:yb-def} to compute the bias generating response variables, and then employing the joint-objective of Equation \ref{eq:jointloss} to compute the final decision boundary between the two class of points. Figure \ref{fig:2d-data} (right) compares this bias-aware decision model with that of a bias-agnostic one (logistic regression with $L_2$ regularization) as shown on the left of the figure.

It can be visualized that the bias-agnostic model is more effective in its decision as evident from a smaller number of erroneous points crossing the decision boundary (shown as the solid line). Yet such a bias-agnostic model, despite being mostly correct, leads to a visible cognitive bias of believing that every point in the upper half of the plot area is green (as per the definitions of sets $P_s$ and $U_{x_2}$). Such biases could lead to detrimental effects (e.g., substitute
`upper-half' with `females' and `green' with `afraid').
On the other hand, the bias-aware model is able to make an adjustment in the decision boundary (comprising piece-wise linear segments) by tilting it away from the dotted horizontal line (defining boundary of $U_{x_2}$). The bias-aware model includes a number of green points below its decision boundary, which as per the earlier example is analogous to assuming that the model in this case predicts that a number of females do not show the negative emotion of fear.

\subsection{Real Dataset}

To demonstrate the potential problems of cognitive biases in prediction systems, for our experiments, we specifically select the task of emotion prediction. Concretely speaking, given a natural language sentence, the task involves predicting the primary emotion of expressed in the sentence from among $5$ possible emotion classes, namely `fear', `anger', `joy', and `sadness' (along with the neutral class).
The dataset that we use in particular for our experiments is the Equity Evaluation Corpus (EEC), compiled by the work in \citep{kiritchenko-mohammad-2018-examining}. In addition to being associated with an emotion, each sentence in this dataset expresses a gender or race. Social identities (gender and race) in \citep{kiritchenko-mohammad-2018-examining} were represented through the use of a set of typical person names (e.g., Ebony is typically a female African American name, whereas Adam is typically a male Caucasian one). The dataset contains an equal proportion of sentences in each different combination of the identity attributes. The sentences were generated by substituting person names and gender specific pronouns from a number of templates expressing different emotions.

A summary of the dataset is presented in Table \ref{tab:DataDesc}.  
\citet{kiritchenko-mohammad-2018-examining} report that even balanced datasets can lead to emotion predictions that are reminiscent of historically prejudiced opinions and cognitive biases, such as: i) women are more prone to be afraid than men, b) African Americans are more prone to be angry etc. The prediction of bias-agnostic models in such cases are affected by the presence of \emph{implicit bias} in the embedded word vectors \citep{gonen-goldberg-2019}.

In addition to experimenting with the balanced dataset, we also prepare a more realistic version of the dataset by 
aligning the emotion distributions to match a set of social stereotypes. The objective of creating these \emph{explicitly biased} datasets was to investigate if our proposed model can reduce biases that are explicitly present as class priors in the data.
One version of such an \emph{explicitly biased} subsampled data represents the stereotype that \emph{males are more prone to anger, whereas women more prone to be afraid} (see the middle part of Table \ref{tab:DataDesc}). Our experiments also revealed that the predictions of bias-agnostic models on the balanced dataset turned out to be biased towards predicting \emph{Caucasians as more likely to be afraid} (possibly due to effects of word embedding on large volumes of text regarding \emph{Islamophobia}). The other subsample that we use for the experiments further aggravates this effect (as can be seen from the bottom part of Table \ref{tab:DataDesc}).

\begin{table}[t]
\centering
\scriptsize
\begin{tabular}{l@{~~}c@{~~}c@{~~}c@{~~}c@{~~}c@{~~}c} 
\toprule
&\multicolumn{5}{c}{Emotion Categories}& \\
\cmidrule(r){2-6}
Identity Attribute &Fear & Anger & Joy & Sadness & Neutral & Total \\
\midrule
Male &1050&	1050&	1050 &	1050&	120	& 4320\\
Female &1050&	1050&	1050 &	1050&	120	& 4320\\
African-American  &700&	700	&700&	700&	120&	2920\\
Caucasian  &700&	700	&700&	700&	120&	2920\\
None  &N/A&	N/A	&N/A&	N/A&	2920 &	2920\\
\midrule
\midrule
\multicolumn{7}{c}{(SS-1): Biased Subsample with (Female, Fear)$\uparrow$, (Male, Anger)$\uparrow$} \\
\midrule
Male &500&	1050&	1050&	1050&	120&	3770\\
Female &1050&	500&	1050&	1050&	120	&3770\\
African-American &450	&550 &	700	&700&	120	&2520\\
Caucasian  &550&	500	&700&700&120&	2570\\
None  &N/A& N/A& N/A & N/A & 1450 & 1450 \\
\midrule
\midrule
\multicolumn{7}{c}{(SS-2): Biased Subsample with (Caucasian, Fear)$\uparrow$} \\
\midrule
Male &850&	1050&	1050&	1050&	120&	3770\\
Female &850&	1050&	1050&	1050&	120	&3770\\
African-American &300	&700 &	700	&700&	120	&2520\\
Caucasian  &700&	700	&700&700&120&	2570\\
None  &N/A& N/A& N/A & N/A & 1450 & 1450 \\
\bottomrule
\end{tabular}
\caption{Distribution of emotion categories (primary task labels) into social identity attributes (gender and ethnicity).}
\label{tab:DataDesc}
\end{table}
%
%

%

\paragraph{Baselines.}
We now describe the details of the different baseline approaches employed in our experiments.

\begin{enumerate}
\item 
\textbf{\STWOR}(\textbf{Bias-Agnostic Learning}): This method employs a degenerate version of the multi-objective learning of Equation \ref{eq:jointloss}, where the only variable used to learn the parameters of the model corresponds to the primary-task labels ($y$'s which in our experiments denote the emotion categories). No bias response variables are used to train model parameters. As word representations, we used pre-trained vectors obtained by applying skip-gram \citep{Mikolov13} on Google-News corpus.
%
\item
\textbf{\STWR}(\textbf{Bias-Agnostic Learning with $L_2$ regularization}): Identical to \STWOR, with the only difference that $L2$ regularization is used to train the model parameters.
The objective was to investigate if the sparsity constraint of $L_2$ regularization (which prevents overfitting) can also help reduce cognitive biases. 

\item
\textbf{\STWRG}(\textbf{Bias-Agnostic Learning with additional Gender Feature}): Since the baselines \STWOR and \STWR do not
use any information from the identity attributes (i.e., gender and race), in this approach we concatenate the gender attribute value (male or female) to each input vector. The objective was to investigate if including social identity specific information can reduce the cognitive biases in the prediction outputs. 

\item
\textbf{\STWRR}(\textbf{Bias-Agnostic Learning with additional Race Feature}): In this approach instead of appending the gender feature, we concatenate the race feature to every input. Again, this We used single objective function with race as an additional feature for learning the primary task here.

\item
\textbf{\STWRDW}(\textbf{Bias-Agnostic Learning with Debiased Word Embedding}): Different from the previous approaches, where we used standard word embedding (which is likely to reflect the cognitive biases in the data), in this baseline, we employ the `neutralize and equalize' debiasing method proposed in \citep{Bolukbasi:2016} to obtain gender-neutral word vectors, e.g. in such an embedded space `babysit' is expected to be equidistant from grandmother and grandfather. The purpose of employing a set of gender neutral word vectors is to reduce the gender stereotypes learned from large volumes of text in the input to a bias-agnostic model and see if neutralized inputs can lead to fairer down-stream predictions. Specifically, we used gender neutralized word embedding obtained by applying transformation on the same set of embedded word vectors that we use for the above baselines (i.e. Google News corpus).      


\end{enumerate}

\paragraph{Variants of bias-aware approaches.}
To compare against the baselines, we train emotion prediction based on our proposed bias-aware multi-objective approach in three different ways, as enumerated below.
\begin{enumerate}
\item \textbf{\MTLG}(\textbf{Bias-aware Multi-objective Learning with Gender only}): In this approach, we learn the prediction model from the multi-objective function of Equation \ref{eq:jointloss} using only the bias variables corresponding to the gender attribute (i.e. $n=1$ in Equation \ref{eq:jointloss}).

\item \textbf{\MTLR}(\textbf{Bias-aware Multi-objective Learning with Race only}): This variant uses a similar approach as above, the difference being this time we use the associations between the ethnicity and the emotion categories to define the biased response generation variables, $y^B$'s.

\item \textbf{\MTLB}(\textbf{Bias-aware Multi-objective Joint Learning}): In this variant, we use both the ethnicity-emotion and the gender-emotion pairs to define two sets of biased response generation variables,
$y^B_1$'s and $y^B_2$'s with $n=2$.
\end{enumerate}

\begin{table*}[t]
\centering
\small
\begin{tabular}{l@{~~}l@{~~}c@{~~~~}c@{~~}c@{~~}c@{~~~~}c@{~~}c@{~~}c@{~~~~}c@{~~}c@{~~}c} 
\toprule
Method & &  &\multicolumn{3}{c}{Fear vs. Gender}  &  \multicolumn{3}{c}{Fear vs. Race} & \multicolumn{3}{c}{Anger vs. Gender}   \\
\cmidrule(r){4-6}
\cmidrule(r){7-9}
\cmidrule(r){10-12}
Name & Sampled & Acc ($A$)  & $\alpha_{\mathrm{female}}$& Fairness & $\gamma$  & $\alpha_{\mathrm{white}}$ & Fairness & $\gamma$ &   $\alpha_{\mathrm{male}}$ & Fairness & $\gamma$  \\
\midrule
\multirow{3}{*}{\STWR} & SS-1 &  0.8337 &1.0000& 0.0000 & 0.0000 &0.3076& 0.2103&0.1679 &0.6187& 0.2169 & 0.1721\\
& SS-2 &  0.7930 &0.9800& 0.0194& 0.0189 &0.2777& 0.1928 &0.1551& 0.6919 & 0.3081 & 0.2218 \\
& None & 0.8620 & 0.7914 & 0.1650 & 0.1384 & 0.2436&0.1842 &0.1517&0.3806 & 0.2357 & 0.1850 \\
\midrule
\STWOR & None &  0.8237 & 0.7971 &0.1617 & 0.1351 & 0.3659& 0.2320 &0.1810 &0.6884 & 0.2145 & 0.1701\\
\STWRDW & None & 0.7380 & 1.0000 &0.0000 & 0.0000 & 0.2636&0.1914 &0.1521&0.7598 & 0.1825 & 0.1463\\
\STWRG & None & 0.7900 & 1.0000 & 0.0000 & 0.0000 & 0.1667&0.1389 & 0.1181 &0.6098& 0.2425 & 0.1855\\
\STWRR & None  &0.7800 &1.0000 & 0.0000 & 0.0000 &0.1877& 0.1524 & 0.1274 &0.6740& 0.2197 & 0.1714\\
\midrule
\MTLG & None & \textbf{0.9430} &\textbf{0.5023} &\textbf{0.2499} & \textbf{0.1986} & 0.2036&0.1540 & 0.1320 & \textbf{0.5817} &\textbf{0.2433} & \textbf{0.1935}\\
\MTLR & None & 0.9400 & 0.7914 &0.1650 & 0.1403 &0.3734& 0.2339 & \textbf{0.1873} &0.6194 &0.2357 & 0.1884\\
\MTLB& None & 0.9100 & 0.5582 &0.1955 & 0.1609 & \textbf{0.3774} & \textbf{0.2349} & 0.1867&0.5940 & 0.2411 & 0.1906\\
\bottomrule
\end{tabular}
\caption{Comparison of emotion classification correctness and fairness (along with a harmonic mean of the two denoted by $\gamma$) for bias-agnostic and bias-aware approaches. Fairness (lack of bias) is measured by associating two emotions classes (fear and anger)
with gender (male/female) and race (black/white). Results show that the bias-aware models output more socially acceptable responses, specifically, a) \emph{not every woman is fearful}, b) \emph{not all Caucasians are phobic}, and c) \emph{not all men are angry}, as evident respectively from the fairness ($F$) values of `Fear vs. Gender', `Fear vs. Race', and `Anger vs. Gender'.}
\label{tab:results}
\end{table*}

\paragraph{Parameters and Settings.}
As seen in Figure \ref{fig:deBiasArch}, the common parameters to all the methods (baselines and proposed) are dimensionality ($d$) of the inputs and that of the shared layer ($p < d$).
We set $d=300$ for all our experiments. All approaches, except \STWRDW, use $300$ dimensional pre-trained skipgram vectors trained on Google News corpus. The vector representation of each sentence is the
sum of embedded representations of constituent words of the sentence.
The inputs in \STWRDW constitute the set of gender-neutralized word vectors \citep{Bolukbasi:2016}.
%
We tuned the dimension of the shared layer in a range of $10$ to $250$ in steps of 10, and report the results only with the optimal value of $p=200$.
%
In all our experiments, we used a train-test split of 80:20. Some emotion-attribute pairs such as `joy vs. gender' exhibit an almost uniform posterior, or in other words, the emotion in these cases are not highly correlated with the identity attribute, e.g. the classifier in this case
does not predict that males are happier than females.
Consequently, we do not employ debiasing on these emotion-attribute pairs.

\paragraph{Evaluation Metrics.}
To address the trade-off between \emph{correct} and \emph{fair} prediction responses, we employ two separate metrics for each. \emph{Correctness} (which we denote as $A$) is measured with the help of accuracy with respect to \emph{all} categorical values of predicted primary task labels, i.e. how many times each input is correctly classified to its ground-truth label.


\emph{Fairness} is computed as a function of the posterior distribution of a \emph{particular} category of primary task (e.g. fear for emotion prediction) with respect to a number of identity attribute values, e.g. (male and female for gender). In particular, for a primary task label $y=l (0\leq l \leq k_s)$ and a binary identity attribute $C=\{U,D\}$ (following the notations of Table \ref{tab:notations}), we compute the product of the posteriors as
\begin{equation}
F=\alpha(1-\alpha),\, \alpha=P(y=l|U),
\end{equation}
which is maximum if $\alpha=\frac{1}{2}$, i.e. when the distribution is uniform. The value of $F$ can thus be used as a fairness measure (higher the better). Note that this argument continues to apply for more than 2 categories.

Since it is desirable to simultaneously obtain a high accuracy and fairness, we combine these two values by taking their harmonic mean to report an overall measure (analogous to measuring F-score from precision and recall). Formally,
\begin{equation}
\gamma = \frac{AF}{A+F},    
\end{equation}
where $A$ is the accuracy with respect to \emph{all} primary task labels, whereas $F$ involves a fairness measure involving \emph{one} such category.

\subsection{Results}

Table \ref{tab:results} reports the results of our experiments. 
First, we observe that on two versions of subsampled datasets (as mentioned in Table \ref{tab:DataDesc}), namely SS-1 (where there is a higher prior of women being associated to fear and men with anger) and SS-2 (with a high prior of whites with fear),
\STWR performs very poorly in terms of the fairness measure (and hence also poorly on the combined metric $\gamma$). Importantly, the results on the subsampled data typically reflects on the fact that predictions under the presence of cognitive biases in data can lead to non-humane responses such as \emph{all women are afraid} (as can be observed from the value $\alpha_{\mathrm{female}}=1$).

The purpose of reporting the results with the sub-sampled (biased) datasets is to demonstrate that the bias in the data is likely to propagate to the predictions. It is also demonstrated that balanced datasets (the \STWR case with no sub-sampling denoted as `None') can somewhat mitigate this bias from the predictions as evident from the increase in $\gamma$ with reference to the SS-1 and SS-2 cases. The use of sub-sampling provides a reference point to compare the effect of adding more annotated data instances to reduce the class imbalance from data (note that in our experiments, the whole dataset is balanced). The disadvantage of adding more data towards balancing the class priors with respect to the identity attributes is that it not only is a manually extensive process requiring retraining the model, but also poses difficulties in foreseeing its effect on the posterior biases.

The observation reported under the `None' column in Table \ref{tab:results} with a $\gamma$ of $0.1384$ (third row) illustrates the important fact that \emph{even a balanced dataset can lead to biased predictions} and this shows that there is further scope for alleviating bias, which is what we explore in the rest of the table.


It can be seen that the use of gender debiased pre-trained word embedding, i.e. \STWRDW does not perform well in terms of reducing gender biases from predictions, our observations in fact corroborates to that of \citep{gonen-goldberg-2019}. The reason this happens is due to the fact that a linear transformation based word embedding debiasing is unable to explicitly take into account the posterior distributions of sensitive combinations of emotion-identity types. Similarly, it also turns out that making use of the gender and ethnicity features as parts of the input data cannot take into account the posterior distributions.

Importantly, we note that the proposed bias-aware methods are
able to reduce three particular cognitive biases, i.e. bias of associating the emotion of fear to women, that of associating anger to men, and that of associating fear to Caucasians, as can be seen by comparing the $\gamma$ values of the bias-aware methods with the bias-agnostic ones. The advantage of the joint method \MTLB over its individual counterparts, i.e. \MTLG and \MTLR is that it reduces the overall $\gamma$ value aggregated across these different biases. From a practical view-point this implies a single predictive model can achieve a trade-off between a given set of specified cognitive biases instead of training separate predictive models to reduce each.

Another important observation is that the overall accuracy values also turn out to be the best among the competing approaches. This happens because the use of pre-trained word embeddings (both gender agnostic and gender equalized) introduce potential sources of gender and race specific biases as parts of the input. Since the ground-truth emotion labels are distributed uniformly across different gender and race categories (see Table \ref{tab:DataDesc}), these biases from large volumes of text contribute to decreasing the effectiveness of the bias-agnostic classifiers. However, with bias-aware training it is possible to make more accurate predictions.

\section{Conclusions and Future Work}

In this paper, we propose a multi-objective learning based framework that seeks to effectively learn to predict a primary task (e.g. emotion classification), with an aim to ensure that such predictions do not constitute social prejudices and stereotypes. More specifically, given a set of identity attributes and a set of sensitive categories (primary task labels), our proposed model seeks to reduce certain pairs of associations that are ethically not correct, e.g. predicting that most black-skinned people are prone to be criminals.
Our experiments on a dataset of emotion prediction shows that this bias-aware learning framework can reduce a number of different cognitive biases from its predictions, such as reducing the number of times the model predicts an emotion of `fear' for a woman etc.


In future, we would like to explore ways of learning to reduce bias without the explicit annotation of social identity specific categorical attributes as parts of the data. More specifically, the idea would be to automatically explore subspaces of data to identify potential candidates of abstract representations of social identities (not necessarily in terms of distinct categories), and associate a set of given sensitive categories to quantify bias estimates against them. The effectiveness of such an approach could then be evaluated by measuring how well the predictions correlate with unprejudiced human judgments.

\textbf{Acknowledgement.}
The first author is supported by Science Foundation Ireland (Grant No. 13/RC/2106).

\bibliographystyle{aaai}
\bibliography{ref}
\end{document}